\documentclass[
 reprint,
 amsmath,amssymb,
 aps,
]{revtex4-2}

\usepackage{graphicx}
\usepackage{dcolumn}
\usepackage{bm}

\usepackage{braket}
\usepackage{color}

\begin{document}

\title{Quantum Computation and Quantum Simulation with Ultracold Molecules}

\author{Simon L. Cornish}
\email{s.l.cornish@durham.ac.uk}
\affiliation{Department of Physics and Joint Quantum Centre Durham-Newcastle, Durham University, South Road, Durham DH1 3LE, UK.}

\author{Michael R. Tarbutt}
\affiliation{Centre for Cold Matter, Blackett Laboratory, Imperial College London, Prince Consort Road, London SW7 2AZ, UK.}

\author{Kaden R. A. Hazzard}
\affiliation{Department of Physics and Astronomy, Rice University, Houston, TX 77005, USA.}

\date{\today}

\begin{abstract}
Ultracold molecules confined in optical lattices or tweezer traps can be used to process quantum information and simulate the behaviour of many-body quantum systems. Molecules offer several advantages for these applications. They have a large set of stable states with strong transitions between them and  long coherence times. They can be prepared in a chosen state with high fidelity, and the state populations can be measured efficiently. They have controllable long-range dipole-dipole interactions that can be used to entangle pairs of molecules and generate interesting many-body states. We review the advances that have been made and the challenges still to overcome, and describe the new ideas that will unlock the full potential of the field.
\end{abstract}

\maketitle

\section{\label{Sec:intro}Introduction}

Producing gases of polar molecules cooled to within a millionth of a degree of absolute zero is now routine in many laboratories, either by associating pairs of alkali-metal atoms~\cite{Ni2008,Takekoshi2014,Molony2014,Park2015,Molony2016,Guo2016,Rvachov2017,Selberg2018,Yang2019,Hu2019,Voges2020,Cairncross2021,Stevenson2022} or by direct laser cooling~\cite{Barry2014,Truppe2017,Anderegg2017, Collopy2018,Vilas2022}. At such ultracold temperatures both the motional and internal states of the molecules can be exquisitely controlled. Moreover, standard techniques from the field of ultracold atoms, such as optical lattices~\cite{Bloch2012} and optical tweezers~\cite{Kaufman2021}, can be used to confine molecules in ordered arrays. These developments unlock many new opportunities in quantum science that exploit the rich properties of molecules.

Molecules, unlike atoms, may possess an intrinsic electric dipole moment allowing the generation of controlled long-range dipole-dipole interactions (DDI) through the application of electric and/or microwave fields. Typical interaction strengths lie between those found in highly magnetic atoms~\cite{Lahaye_2009,Chomaz2022} and Rydberg atoms~\cite{Saffman_2010,Browaeys2020}, providing a different regime for study. Additionally, vibration and rotation of the molecule lead to a rich internal structure of long-lived states.  This combination of properties leads to new experimental possibilities and a broad range of applications, spanning ultracold chemistry \cite{Krems2008}, few-body physics \cite{Quemener2012}, precision measurement \cite{Hudson2011}, quantum simulation \cite{baranov2012condensed} and quantum computation \cite{DeMille2002}.

In this review we focus on the use of ultracold molecules for quantum simulation and quantum computation. We outline the new opportunities in these fields stemming from the properties of molecules. We give a snapshot of the current experimental state of the art and describe the challenges specific to molecules.  We present a vision for the immediate next steps in the field and describe several nascent ideas which could unlock the full potential of ultracold molecules.

The interested reader is directed to earlier reviews~\cite{Carr2009,Lemeshko2013,Wall2015b,Bohn2017,Koch2019,softley2023cold} that chart the remarkable progress in the field.

\begin{figure*}
    \centering
    \includegraphics[width=2\columnwidth]{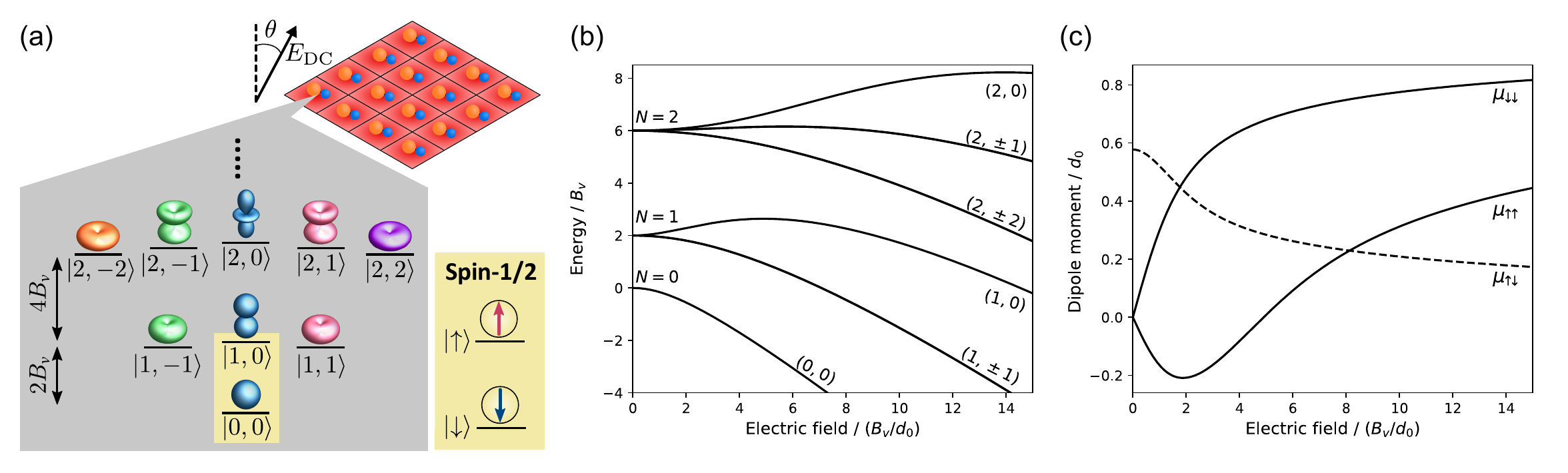}
    \caption{(a) Pseudo-spins (or qubits) may be encoded in the rotational states of ultracold polar molecules confined in optical lattices (or tweezer arrays) for applications in quantum simulation and  computation. Typical rotational constants, $B_v$, lead to transitions between rotational states in the microwave domain. Spin-exchange and Ising interactions between molecules may be controlled through the applied field, $E_{\rm{DC}}$. (b)~Energy and (c)~dipole moments of a rigid rotor in an electric field. Energies are labelled by $(N,M_N)$ at high field. Dipole moments are shown for $\ket{\downarrow}=\ket{0,0}$ and $\ket{\uparrow}=\ket{1,0}$. All  quantities are scaled to the rotational constant $B_v$ and molecule-frame dipole moment $d_0$. (c) The tunability of the spin-exchange  ($\propto \mu_{\uparrow\downarrow}^2$) and  Ising interaction ($\propto \left( \mu_{\uparrow\uparrow} - \mu_{\downarrow\downarrow}
 \right)^2$).  (a) Adapted from~\cite{Blackmore2018} with permission from the authors.
        }
    \label{fig:theory}
\end{figure*}

\section{\label{Sec:theory}Theoretical considerations}

We  briefly review the basic theory of molecular structure and interactions that motivates the growing interest in ultracold molecules. We explicitly considers a  $^1\Sigma$ diatomic molecule in its vibrational ground state, but the long-ranged  coupling of rotational states holds more generally~\cite{brown2003rotational}.   Such molecules are described by a rigid-rotor Hamiltonian with eigenstates   $\ket{N,M_N}$ labeled by the rotational angular momentum  $N$ and its projection $M_N$ along the quantisation axis. The rotational state energies form a ladder with anharmonic spacing given by $E_N=B_vN(N+1)$, where $B_v$ is the rotational constant (see Fig.~\ref{fig:theory}(a)). Transitions between neighboring rotational states typically lie in the microwave domain. The  rotational levels are $(2N+1)-$fold degenerate, but in practise, there is hyperfine structure and the degeneracy is lifted by an applied magnetic field.  We therefore can consider  $\ket{N,M_N}$ states as  isolated, and a pair of states may  encode a spin-1/2 or qubit.

The symmetry of the $\ket{N,M_N}$ states leads to vanishing electric dipole moments in the laboratory frame. However, controllable DDI can be engineered in several ways. An  electric field can be used to create static dipole moments (see Fig.~\ref{fig:theory}(b)\,\&\,(c)). Microwave fields can be used to create rotational-state superpositions with oscillating dipole moments.  Finally a pair of molecules can be prepared in neighboring rotational states connected by an allowed electric-dipole transition. The DDI between molecules $i$ and $j$ with dipole moments $\vec{\mu}_{k}$ separated by   $\vec{r}_{ij}=\vec{r}_i-\vec{r}_j=r_{ij}\vec{e}_{ij}$ is 
\begin{align}
V_{ij}^{\mathrm{DDI}}
 &=
 \frac{1}{4\pi\epsilon_0}
 \frac{ \vec{\mu}_i \cdot \vec{\mu}_j
       - (\vec{\mu}_i \cdot\vec{e}_{ij}) (\vec{\mu}_j \cdot\vec{e}_{ij}) }{r_{ij}^3} \:.
 \label{Eq:DDI}
\end{align}
In Sections~\ref{Sec:lattice}\,\&\,\ref{Sec:tweezers} below, we present  how the interaction in Eqn.~(\ref{Eq:DDI}) can be mapped onto a subset of states to encode a spin or qubit and perform quantum simulations and quantum information processing. We will explain how the choice of states and applied fields dictates the nature of  the interactions, and how some  non-interacting state combinations can be useful for storing quantum information. Throughout it is useful to keep in mind that typical electric dipole moments of diatomic molecules are $\sim1$\,Debye resulting in interaction strengths $\sim1$\,kHz for molecules spaced by 532\,nm in an optical lattice.

\section{\label{Sec:lattice}Molecules in lattices for quantum simulation}

\subsection{Model Hamiltonians and many-body phases}

Quantum simulations with ultracold molecules divide into two classes, one where molecules are frozen in place by a lattice or microtraps to eliminate losses, and one where the molecules can move and losses are tolerated or mitigated. In the former,  frozen motion  necessitates using rotational states to have interesting many-body physics and  generally leads to  interacting spin models as studied in  quantum magnetism. In the latter,  interesting physics can occur already with the rotational  ground state.  

When molecules are frozen in place,  projecting   Eqn.~\eqref{Eq:DDI} onto the    rotational states  populated experimentally gives a spin model. With  two rotational states,  a unit-filling array  
is described by the dipolar XXZ model,
 \begin{align}
 H_\mathrm{XXZ}
 &= \frac{1}{2}\sum_{i\neq j}\left[ \frac{J_\perp}{2} \left(S_i^+ S_j^- + S_i^- S_j^+\right)
  + J_z S_i^z S_j^z \right] 
 \label{eq:XXZ}
\end{align}
where the $\{S_j^a\}$ are the site-$j$ spin operators.
Refs.~\cite{Gorshkov2011,Gorshkov2011b} theoretically showed  how  experiments could engineer  Eqn.~\eqref{eq:XXZ}, building on a body of research starting in Ref.~\cite{Barnett2006} and  reviewed in Ref.~\cite{Wall2015b}.
  $J_\perp$ and $J_z$ depend on the  states chosen and the electric field,  illustrated in Fig.~\ref{fig:theory}. As an example,  internal states $\ket{\downarrow}=\ket{N=0,M_N=0}$ and $\ket{\uparrow}=\ket{N=1,M_N=0}$ yield
\begin{align}
 J_\perp
 &=\frac{1-3\cos^2\theta_{ij}}{4\pi\epsilon_0 
 r_{ij}^3}
 \times
 2\mu_{\uparrow\downarrow}^2 \:, \\
 J_z
 &=\frac{1-3\cos^2\theta_{ij}}{4\pi\epsilon_0 r_{ij}^3}
 \times\left( \mu_{\uparrow\uparrow} - \mu_{\downarrow\downarrow}
 \right)^2,
 \label{eq:Js}
\end{align}
where $\mu_{\sigma\tau}=\braket{\sigma|{\hat\mu}^{(z)}|\tau}$  and ${\hat \mu}^{(z)}$ is the $z$-component of ${\hat \mu}_i$.
The $J_\perp$ term describes spin-exchange;  the $J_z$ term describes Ising interactions, which vanish absent an electric field. Both prefactors depend on the angle $\theta_{ij}$ between the dipole moments and the intermolecular vector, highlighting the  anisotropy  of the DDI: side-by-side  dipoles repel, while head-to-tail dipoles attract. The form of the interactions is the same for other state choices, but  magnitudes change~\cite{Wall2015b,Gorshkov2011b}. More generally, below unit filling,   Eqn.~\eqref{Eq:DDI} gives two additional interactions:  spin-density  [$W(S_i^z n_j + n_iS_j^z)]$ and  density-density  [$V n_i n_j$], where $n_i$ is the site-$i$ number operator. Allowing tunneling results in a highly-tunable $t$-$J$-$V$-$W$ model~\cite{Gorshkov2011,Gorshkov2011b}.

The  model Eqn.~\eqref{eq:XXZ} is a long-ranged  interacting variant of the widely-studied XXZ model. It generalizes  Ising  ($J_\perp=0$), XY ($J_z=0$), and Heisenberg ($J_\perp=J_z$) models that are   studied intensely in statistical mechanics, condensed matter, and many-body physics~\cite{auerbach1998interacting,sachdev2008quantum,lacroix2011introduction}. Their ground states  display a  broad range of symmetry breaking phases, topological phases, including, spin liquids~\cite{Yao2018}, and quantum phase transitions  depending on $J_\perp/J_z$ and the   lattice geometry. Driven out of equilibrium,  integrable dynamics, integrability breaking, and   unconventional superdiffusive scaling   in one dimensional systems have  been predicted~\cite{vznidarivc2011spin,ljubotina2019kardar,gopalakrishnan2019kinetic,de2019anomalous}.

Several theoretical proposals give blueprints for spin-1/2 and higher-spin models  beyond the ``naturally-occurring" XXZ model. The most common  techniques are   microwave dressing   the rotational states~\cite{Gorshkov2011}   and  Floquet engineering  with periodic microwave pulse sequences~\cite{choi2020robust,geier2021floquet,scholl2022microwave}, as recently demonstrated experimentally~\cite{Christakis2023}. Both   allow flexible control of  the two-body interaction -- sums of  $S_i^\alpha S_j^\beta$  and significant spatial anisotropy can be created. These can engineer spin models with interacting symmetry-protected topological ground states~\cite{manmana:topological-phases_2012}, spin-orbit coupling~\cite{Syzranov2014} (related physics has been seen in Rydberg atoms~\cite{Lienhard2020}), and the famous Kitaev honeycomb model that  harbors exotic non-Abelian topological order~\cite{kitaev2006anyons,gorshkov:kitaev_2013}  and can robustly encode and process quantum information.  
 Raman lasers can induce spatially-varying  interactions and exotic    fractional Chern~\cite{Yao2013} and Hopf insulators~\cite{Schuster2021}. 
 Recent work discussed in Sec.~\ref{sec:beyond-qubits}  has explored the use of more rotational states to realise synthetic dimensions~\cite{ozawa2019topological,Sundar2018,hazzard2023synthetic}.  We note that  earlier influential work suggested an alternative  implementation of spin models with ultracold molecules~\cite{Micheli2006,brennen2007designing}. Although extremely flexible,  this approach requires ${^2}\Sigma_{1/2}$ molecules and has yet to be attempted.

The molecules need not be pinned to a lattice site, or confined in a lattice at all, and this opens up quantum simulations of itinerant particles. However, this allows fast inelastic losses to occur, even in molecules that are ostensibly non-reactive  (see also the Perspective Ref.~\cite{Langen2023}  in this issue), which must be suppressed or circumvented. In a lattice, the continuous quantum Zeno effect can suppress losses: if molecules start on different sites, they are  forbidden from hopping onto the same site by  continuous strong ``measurements" associated with  the on-site reactive losses, and long lifetimes are  experimentally observed~\cite{Yan2013,zhu:suppressing_2014}. In weaker lattices or bulk gases, dc electric-field shielding~\cite{Valtolina2020,matsuda2020resonant,Li2021} or microwave shielding of reactions~\cite{gonzalez2017adimensional,karman2018microwave,lassabliere2018controlling,anderegg2021observation,Schindewolf2022,Chen2023,bigagli2023collisionally} can suppress losses by adding a  repulsive barrier to the short-ranged interaction potential. 

If reactive losses can be successfully suppressed, itinerant models  offer a wealth of  important physics. The $t$-$J$-$V$-$W$ model mentioned above 
can occur in either bosonic or fermionic variants. The fermionic  version is an extension of the usual $t$-$J$ model, which in certain limits approximates the Hubbard model~\cite{fazekas1999lecture}. The Hubbard model is central to condensed-matter physics, in part  as a minimal model  capturing the interplay of tunneling  and interactions  and in part for its relation to  high-temperature superconducting cuprates~\cite{qin2022hubbard,arovas2022hubbard,bohrdt2021exploration}. The $t$-$J_\perp$ model arising without an electric field was predicted in Ref.~\cite{manmana:correlations_2017} to have several ground-state  phases even in one dimension, including spin-density waves, charge-density waves, spin-gapped superconductors, and phase separated regimes, and correlations whose long-distance power-law decay    relates non-trivially   to the dipolar interaction power law, and Ref.~\cite{kuns:d-wave_2011} predicted $d$-wave superfluids on a checkerboard. The two-dimensional case is numerically challenging and  an excellent quantum simulation target, with important open questions, e.g. how holes bind.

Itinerant physics requires neither a lattice nor the use of multiple rotational states; molecules in the bulk in their rotational ground state are expected to show  interesting phases of matter, including  $p$-wave topological superfluids in two-dimensional fermionic molecules~\cite{baranov:superfluid_2002,Cooper2009,shi:singlet_2010}, superfluid liquid crystals~\cite{wu2016liquid},  supersolids in self-bound droplets~\cite{schmidt:self-bound_2022}, and Wigner crystals when interactions are dressed~\cite{buchler2007strongly,micheli:cold_2007, rabl2007molecular}.  With a lattice, but using only rotational ground states, extended Hubbard Hamiltonians can be realized, which possess a variety of ordered patterns with and without superfluidity~\cite{Capogrosso-Sansone2010,baranov2012condensed}. 

While itinerant quantum simulations potentially display a wide variety of phenomena, earlier 
theory neglected both  reactive losses and any shielding to suppress them. 
The detailed structure of the interaction potential will differ when shielding is present, requiring modification of previous theory. 
While this may change details, many phenomena are likely to survive.

\subsection{Experimental and theoretical challenges}

Many quantum simulations discussed above require molecules in coherent rotational-state superpositions in lattices with   fillings approaching one molecule per site, and in the lowest energy band for  experiments where tunneling plays a role. This is  challenging for several  reasons. 

Molecules exhibit an anisotropic polarisability where, e.g. for a diatomic molecule, the polarisability along the bond axis generally differs from that orthogonal to the bond. This leads to 
light shifts that depend on $N$, $M_N$, and the angle between the laser polarisation  and the quantisation axis~\cite{Kotochigova2010,Neyenhuis2012}. Further complications from  hyperfine states result in a rich, trap intensity-dependent level structure~\cite{Gregory2017,Blackmore2018}. The resulting differential light shifts severely limit the bare single-molecule coherence times and can mask the (typically $\sim0.1-10$\,kHz) DDI.

Notably, achieving high filling of molecules in a lattice remains challenging. An early   proposal~\cite{Damski2003} suggested  dual-species atomic Mott insulators~\cite{Jaksch1998,Greiner2002} could produce one atom of each species per  site, with  atom pairs subsequently  converted to molecules.   In practise this approach is difficult: it is  sensitive to the species-dependence of the  interactions,  spatial overlaps, and the relative atomic polarisabilities.  
A modified  protocol using superfluid $^{87}$Rb and Mott insulating $^{133}$Cs with tuned  interactions achieved a  filling  exceeding 30\,\% for Feshbach molecules~\cite{Reichsollner2017}. Similar fillings  have been reported for ground state $^{40}$K$^{87}$Rb  by combining a Mott insulator of $^{87}$Rb   with a band insulator of $^{40}$K~\cite{Moses2015}. Although these  allow access to significant many-body physics,  higher filling is desirable.  The recent creation of quantum degenerate molecular gases (Section~\ref{Sec:outlook}A)  provides a promising path to  low-entropy molecules in lattices.

A major additional challenge for quantum simulation with molecules is suppressing  losses  to explore the many-body phases in itinerant molecules, though shielding techniques may alleviate this. Other  frontiers include  
implementing  single-site addressing and entropy engineering, realising  more complex Hamiltonians, and extending to new ultracold molecules, including those produced by direct laser cooling. 

Finally,  calculating the properties of molecular systems is often extremely difficult, whether one- and two-molecule properties,  such as susceptibilities and scattering cross sections, or many-body behaviors. 
This  causes issues for quantum simulation  experiments: difficulty preparing low-temperature equilibrium states, measuring their temperature, or diagnosing their validity. However, progress is being made  even where properties cannot be calculated, for example   fluctuation-dissipation relations  for thermometry~\cite{zhou2011universal,hartke_thermometry_2020,pasqualetti2023equation},  Lieb-Robinson bounds to guarantee stability of local observables under small imperfections~\cite{trivedi2022quantum}, and quantum control for state preparation~\cite{muller2022one}. Although these techniques need to be adapted to molecules,  their general principles  apply.
These challenges are also  a blessing: the difficulty predicting the behaviour provides an opportunity for experiments to act as powerful quantum simulators that give insights into physics that cannot be predicted classically.  

\subsection{Measurements of spin dynamics}

\begin{figure*}
    \centering
    \includegraphics[width=2\columnwidth]{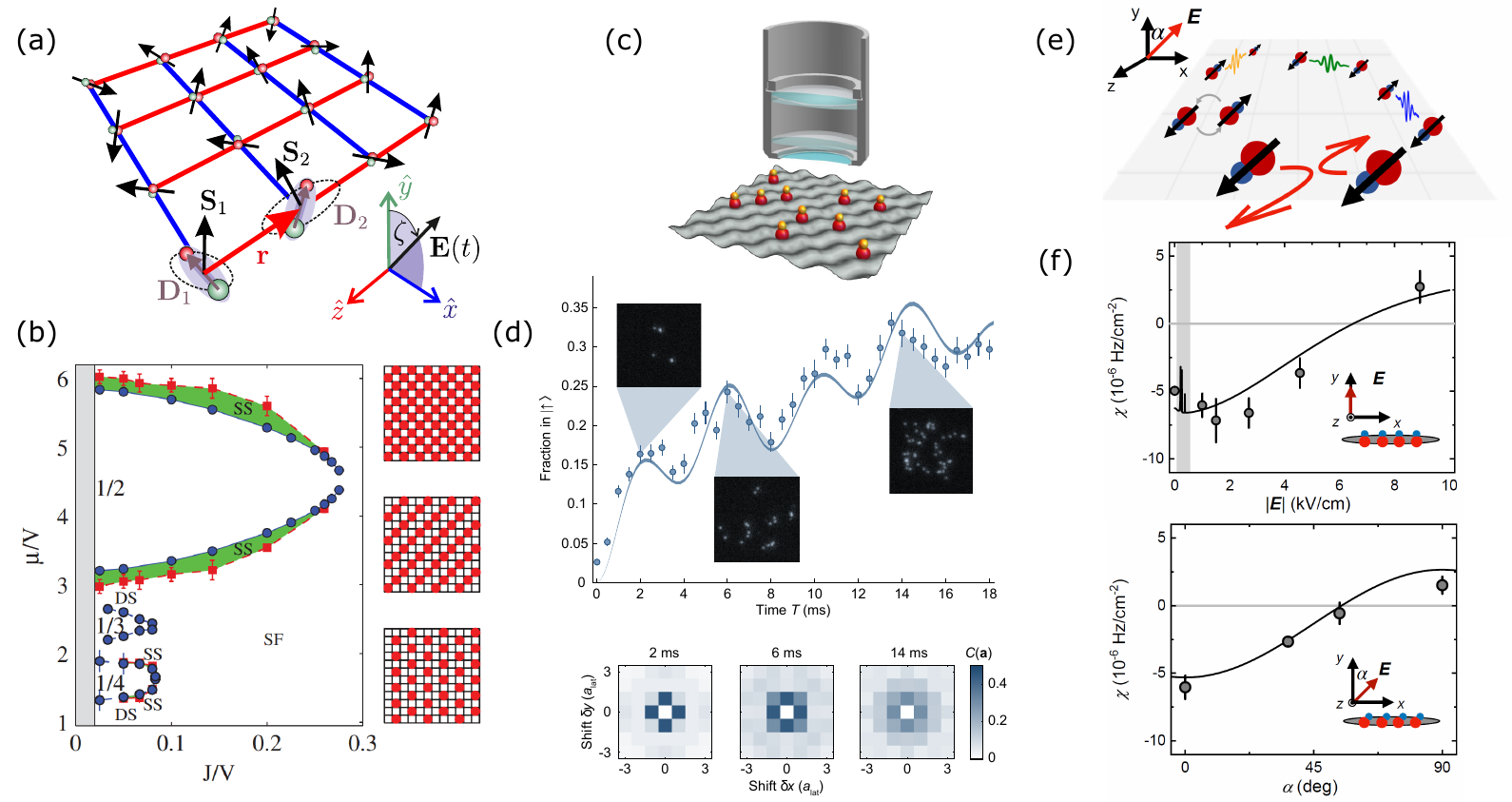}
    \caption{Quantum simulation with ultracold polar molecules. (a) Molecules pinned to the sites of an optical lattice can be used to engineer  spin models.  (b) Novel quantum phases become accessible when the molecules are able to tunnel in the lattice. The zero-temperature phase diagram as a function of tunnelling rate $J$ and chemical potential $\mu$ (both scaled to the DDI strength, $V$) exhibits a devil's staircase (DS) with checkerboard, stripe and star Mott lobes (shown right), as well as superfluid (SF) and supersolid (SS) phases. Other situations show topological phases.  (c) Schematic of a molecular quantum gas microscope used to image individual $^{23}$Na$^{87}$Rb molecules in a 2D lattice with single-site resolution. (d) Observing correlations between molecules interacting in an XY spin model. The insets show typical fluorescence images and, below, their associated correlation matrices. (e) Exploration of itinerant quantum magnetism in a thermal 2D gas of $^{40}$K$^{87}$Rb molecules. (f) Controlling the spin-spin interaction ($\chi$) with the strength (top) and direction (bottom) of the applied electric field. (a) Adapted from~\cite{Micheli2006}, (b) adapted from~\cite{Capogrosso-Sansone2010}, (c)-(d) adapted from~\cite{Christakis2023} and (e)-(f) adapted from~\cite{Li2023}, with permissions from the authors.}
    \label{fig:spinexchange}
\end{figure*}

The field of  quantum simulation of spin models with ultracold molecules has just begun, with three experiments beginning to show its power.
The spin-exchange interactions ($J_\perp$ term in Eqn.~\eqref{eq:XXZ}) were first observed using $^{40}$K$^{87}$Rb molecules in a three-dimensional optical lattice~\cite{Yan2013} following a technique proposed in Ref.~\cite{Hazzard2013}. The molecules were initialised in   $\ket{\downarrow}=\ket{0,0}$ and interrogated with a Ramsey sequence on a  $\sim2.2$\,GHz microwave transition to  $\ket{\uparrow}=\ket{1,-1}$. Experiments observed that the Ramsey contrast oscillated as a function of the time between    $\pi/2$ pulses  with a frequency  matching the expected nearest-neighbor interaction strength, $J_{\perp}/2\simeq h \times 52$~Hz, and decayed with a dependence on density characteristic of DDI~\cite{Yan2013};   Ref.~\cite{Hazzard2014}  used $\ket{\uparrow}=\ket{1,0}$ and observed the expected change of DDI strength. Ref.~\cite{Yan2013} also  allowed the molecules to tunnel along one direction and observed  the continuous quantum Zeno effect induced by fast on-site loss rates. These experiments showed that even  lattice fillings of 10-25\,\% and bulk measurements can observe DDI effects, laying the foundations for the study of  many-body dynamics in disordered dipolar systems. 

Recent work used a quantum gas microscope  to measure site-resolved spin dynamics  of $^{23}$Na$^{87}$Rb in a two-dimensional lattice, again with  two rotational states as an effective spin-1/2~\cite{Christakis2023}. This builds on  the  observation of the Hanbury Brown-Twiss effect in molecules, the first application of a molecular quantum gas microscope~\cite{Rosenberg2022}.  
Spin correlations were measured  (Fig.~\ref{fig:spinexchange}(c) and (d)), isolating first, second, and third-nearest neighbors which, combined with varying  the quantisation axis,  directly evidenced the DDI's anisotropy. They also engineered the XXZ Hamiltonian starting from the $J_z=0$  by using a Floquet technique~\cite{Geier2021,Scholl2022} where a microwave pulse sequence creates the desired  time-averaged Hamiltonian. Quantum gas microscopy of  ultracold polar molecules is a pivotal result with great promise for the future.

Li et al. have  studied dynamics involving both spin and motion~\cite{Li2023}. They combine DC electric and AC microwave fields to control $^{40}$K$^{87}$Rb in isolated 2D planes (see Fig.~\ref{fig:spinexchange}(d)). Their  DC electric field  gives $J_z \ne 0$  in Eqn.~\eqref{eq:XXZ}, and the spin dynamics  depend on  $\chi=J_z-J_\perp$. 
The DDI slightly shifts  the rotational transition frequency   ($\sim100$\,Hz), which is  measured using a Ramsey sequence with XY8 dynamical decoupling   to achieve sufficiently long coherence times.  Their measurements (Fig.~\ref{fig:spinexchange}(f)) demonstrate  control of the interactions using  both the strength and orientation of the  electric field. The field orientation affects the spin-exchange and Ising interactions similarly  (Eqn.~\eqref{eq:Js}), whereas its magnitude  controls their relative strength (see Fig.~\ref{fig:theory}\,(c)). At short times the molecules move little so  dynamics are primarily of the spins, and switching the internal state of the molecule can coherently reverse the spin dynamics. At longer times, collisions  irreversibly decohere and  thermalise the molecules~\cite{Valtolina2020}. Stronger DDI was observed to lead to faster  collisions  and, hence,  faster coherence loss. The authors point to future studies of the Heisenberg  model ($\chi=0$)  by tuning the electric field to $\sim7$\,kV/cm, as shown in Fig.~\ref{fig:spinexchange}(f). Collectively the results highlight the  tunability of spin Hamiltonians implemented using ultracold polar molecules. 

Importantly, much of the  recent progress stems from improved control and understanding of molecular quantum states in applied fields, as discussed in a Perspective in this issue~\cite{Langen2023}. Methods to eliminate or suppress differential light shifts of rotational transitions in optical potentials, thereby extending coherence times, have been crucial. 
Often the intrinsic single-molecule coherence time is only $\sim1$\,ms~\cite{Blackmore2018,Li2023}. Methods to match the polarisabilities of different rotational states have included tuning the trap polarisation~\cite{Kotochigova2010, Neyenhuis2012, Seesselberg2018, Burchesky2021, Tobias2022}, intensity~\cite{Blackmore2018} and ellipticity~\cite{Park2023}. Recently magic-wavelength trapping~\cite{ye:quantum_2008,derevianko:colloquium_2011} has emerged as an effective solution~\cite{kotochigova:controlling_2006,zelevinsky:precision_2008,  Guan2021,bause2020}. Indeed, the experiments using $^{23}$Na$^{87}$Rb molecules discussed above used a near-magic wavelength lattice where the polarisabilities of the two relevant rotational states differed by less than 1\,\%, leading to bare single-molecule coherence times of 56(2)\,ms. In very recent work, second-scale rotational coherences have been reported for $^{87}$Rb$^{133}$Cs molecules, achieved by tuning the trap wavelength near weakly-allowed transitions to the   $\mathrm{b}^3\Pi-\mathrm{A}^1\Sigma^+$ potential~\cite{Gregory2023}. At the same time, more elaborate microwave pulse sequences, such as  XY8 dynamical decoupling, have been used~\cite{Holland2023, Bao2023,Li2023} to reduce single-particle
dephasing; Ref.~\cite{Li2023}  improved the coherence time by a factor of $\sim70$ from 0.24(1) to 17(1)\,ms. With these techniques, rotational coherences have become a usable resource and further quantum simulation experiments  are eagerly anticipated.

\section{\label{Sec:tweezers}Molecules in tweezers for quantum computation}

Molecules provide many stable hyperfine and rotational states that make attractive qubits, and long-range interactions suitable for entangling operations. Fast single qubit gates can be driven using microwave fields or two-photon optical processes. State-selective detection can be achieved by driving an optical cycling transition in the molecule or, following coherent dissociation, in one of the constituent atoms.

\subsection{Tweezer arrays}

An array of optical tweezer traps, each containing a single molecule, is an ideal starting point for quantum computation with molecules. In 2019, Anderegg et al.~\cite{Anderegg2019} trapped laser-cooled CaF molecules in an array of five tweezers. The molecules were cooled from a MOT into a dipole trap using $\Lambda$-enhanced gray molasses cooling to increase the density, then loaded into tweezers and detected with high fidelity by fluorescence imaging.  Light-assisted collisions result in a collisional blockade, ensuring zero or one molecule per tweezer. These methods have now been extended to larger arrays and tighter confinement, enabling entanglement of pairs as described below. 

In 2021, Cairncross et al.~\cite{Cairncross2021} created single ground-state NaCs molecules in optical tweezers. First, Na and Cs were loaded into separate tweezers from a dual species MOT.  Fluorescence imaging was used to select experiments containing an atom pair. These atoms were then cooled to the motional ground state by polarization gradient cooling followed by Raman sideband cooling, and subsequently prepared in single quantum states. Then, the traps were merged adiabatically and the pair converted into a weakly-bound molecule by sweeping a magnetic field through a Feshbach resonance. Finally, the weakly-bound molecule was transferred to a single hyperfine component of the rovibrational ground state by  a two-photon Raman transition. The efficiency of producing the ground state molecule from a pair of atoms was $31 \pm 4$\%, the probability of occupying the motional ground state was $65 \pm 5$\%, and the lifetime of the ground-state molecule in the tweezer was $3.4 \pm 1.6$~s. This pioneering work was soon extended to small 1D arrays~\cite{Zhang2022}. Similar results have now been reported for RbCs molecules~\cite{Guttridge2023}. 

\subsection{Qubits and coherence times}

High-fidelity gate operations require  a coherence time much longer than the gate time. For molecules,  long hyperfine and rotational coherence times require a qubit that has low sensitivity to fluctuations in magnetic field and tweezer trap intensity. The qubit transition should also be well separated from other transitions so that single-qubit gates can be driven quickly without coupling to any other states. This can be   challenging for molecules with large nuclear spins and small rotational constants, where the density of states is high. 

In their ground-states, bialkali molecules have no electronic angular momentum so the hyperfine structure originates exclusively from the nuclear spins and is highly insensitive to external fields. Nuclear spin superpositions  can be prepared by driving a two-photon transition. Using NaK molecules in an optical dipole trap, Park et al.~\cite{Park2021} demonstrated a hyperfine coherence time close to one second. To go further, Gregory et al.~\cite{Gregory2021} identified a pair of nuclear spin states in the ground state of RbCs that have identical magnetic moments at a specific magnetic field. At this field the magnetically-insensitive qubit has a long coherence time, limited by a tiny differential ac Stark shift coupled with the non-uniform intensity of the trapping light. For light linearly polarized at angle $\theta$ to the magnetic field, this tensor shift is proportional to the Legendre polynomial $P_2(\cos\theta)$. Setting $\cos\theta = 1/\sqrt{3}$ thus eliminates the differential Stark shift. In this case, no decoherence  of the hyperfine qubit  was discernible and a coherence time exceeding 5.6~s was deduced.

Rotational coherence times have been studied for CaF  in tweezer traps~\cite{Burchesky2021}. Here, a pair of states in neighboring rotational manifolds forms a qubit that, to first order, is magnetically insensitive around zero field. Although there is a differential light shift, its gradient with intensity, at some chosen intensity, can be tuned to zero through the choice of polarization angle. Combined with active cancellation of magnetic field noise, the experiment measured a rotational coherence time of $93 \pm 7$~ms for molecules cooled to 5~$\mu$K. A spin echo extended the coherence time  to $470 \pm 40$~ms, probably limited by residual sensitivity to the intensity variatios across the trap.

\subsection{Entanglement}

DDI can be used to entangle molecules and  implement two-qubit gates~\cite{DeMille2002,Yelin2006,Zhu2013,Ni2018,Hughes2020}. The first paper on this topic~\cite{DeMille2002} proposed a 1D array of molecules polarised by an electric field, $E_{\rm{DC}}$, with  qubits $\ket{\downarrow}=\ket{\tilde{N}=0,M_N=0}$ and $\ket{\uparrow}=\ket{1,0}$, whose dipole moments depend on  $E_{\rm{DC}}$ (Fig.~\ref{fig:theory}). A small electric field gradient along the array shifts the qubit frequency to provide single-site addressability. The two-qubit gate uses the Ising interaction from Eqn. (\ref{eq:XXZ}), proportional to $\left( \mu_{\uparrow\uparrow} - \mu_{\downarrow\downarrow}
 \right)^2$, to distinguish the frequencies of the molecule-pair transitions $\ket{\downarrow}\ket{\downarrow} \leftrightarrow \ket{\downarrow}\ket{\uparrow}$ and $\ket{\uparrow}\ket{\downarrow} \leftrightarrow \ket{\uparrow}\ket{\uparrow}$. A microwave or two-photon pulse that resolves this frequency difference implements a two-qubit gate. 
 
Alternatively, molecules can be entangled without applying $E_{\rm{DC}}$ by using the spin-exchange interaction of Eqn.~(\ref{eq:XXZ}), proportional to $\mu_{\uparrow\downarrow}^2$. A molecule pair prepared in $\ket{\downarrow}\ket{\uparrow}$ will evolve into $\ket{\uparrow}\ket{\downarrow}$. Reference~\cite{Ni2018} shows one way of using this spin-exchange for two-qubit gates. Here, the qubit states $\ket{0}$ and $\ket{1}$ are hyperfine levels of the ground rotational state, and a rotationally excited state $\ket{e}$ is used for the interaction. A pair of molecules from an array is moved to an interaction zone, a microwave $\pi$-pulse transfers $\ket{1} \rightarrow \ket{e}$, the pair are brought close together and then separated again so that the time-integrated interaction swaps $\ket{1}\ket{e}$ and $\ket{e}\ket{1}$, and then a second $\pi$-pulse transfers $\ket{e} \rightarrow \ket{1}$ to complete the gate. All other molecules have a different tweezer intensity to the selected pair, so are light-shifted out of resonance with the microwave field. 

A similar method uses stationary molecules in stationary states. In the $\ket{\downarrow}$, $\ket{\uparrow}$ representation, the eigenstates of the DDI are $\ket{\downarrow}\ket{\downarrow}$, $\ket{\uparrow}\ket{\uparrow}$ and $\ket{\Psi^{\pm}}=\frac{1}{\sqrt{2}}(\ket{\downarrow}\ket{\uparrow} \pm \ket{\uparrow}\ket{\downarrow})$. The   entangled states $\ket{\Psi^{\pm}}$ are separated by the interaction energy, $2J_{\perp}$. A $2\pi$-pulse resonant with  $\ket{0}\ket{0} \leftrightarrow \ket{\Psi^+}$ implements a two-qubit gate. In practice, it is better to use qubit states with no electric dipole coupling and an auxiliary state for the two-qubit gate. A simple implementation is described in Ref.~\cite{Caldwell2020c}. A more sophisticated one uses an optimized microwave field  pulse to produce a gate that is robust to errors in the trapping and control fields~\cite{Hughes2020}, and  gate fidelities exceeding 99.9\% are predicted to be feasible. 

\begin{figure*}
    \centering
    \includegraphics[width=2\columnwidth]{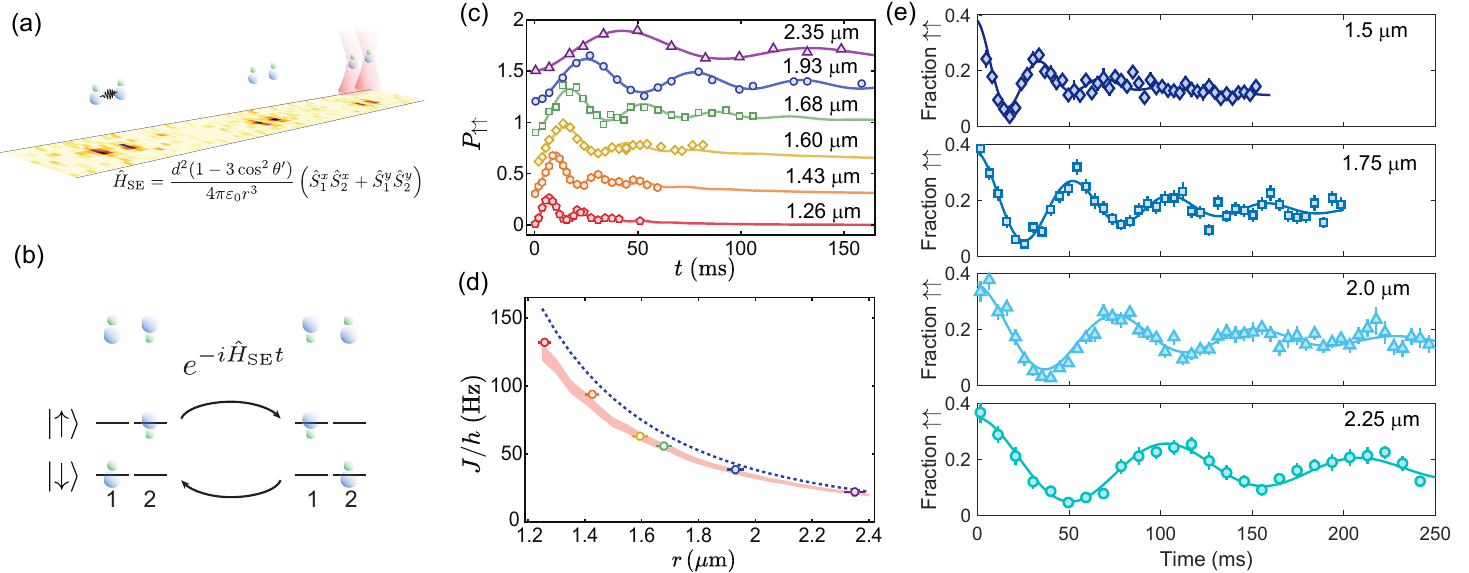}
    \caption{Entangling gate between CaF molecules  using dipolar spin-exchange interactions. (a) Single molecules in a tweezer array with dipolar interactions $H_{\rm SE}$. (b) Exchange of rotational excitation induced by $H_{\rm SE}$. (c) Spin-exchange oscillations at various tweezer separations, from \cite{Holland2023}. (d) Points: measured interaction strength  versus separation; Dashed line: zero-temperature theory; Red band: theory at the finite temperature of the molecules. (e) Spin-exchange oscillations at various tweezer separations, from \cite{Bao2023}. (a)-(d) adapted from \cite{Holland2023}, (e) adapted from \cite{Bao2023}, with permissions from the authors.}
    \label{fig:gate}
\end{figure*}

Two remarkable papers~\cite{Holland2023, Bao2023} have demonstrated deterministic entanglement of molecule pairs using dipolar spin-exchange interactions.  Both used  rotational qubits of CaF trapped in tweezer arrays (Fig.~\ref{fig:gate}(a)). 
After bringing molecule pairs initialized in $\ket{\uparrow\uparrow}$ close, the experiments applied a  Ramsey sequence -- a pair of $\pi/2$ pulses separated by an evolution period $t$ -- and a dynamical decoupling sequence  to extend coherence times, and then measured the population in $\ket{\uparrow\uparrow}$, $P_{\uparrow\uparrow}$. Spin-exchange causes $P_{\uparrow\uparrow}$ to oscillate with an angular frequency of $J_{\perp}/(2 \hbar)$. Figure \ref{fig:gate}(c),   from Ref.~\cite{Holland2023}, shows these   oscillations. The measured value of $J_{\perp}$ as a function of molecule separation $r$ (Fig.~\ref{fig:gate}(d))   is found to follow the theoretical prediction once the temperature of the molecules is accounted for. Figure \ref{fig:gate}(e) shows another set of spin-exchange oscillations,  from Ref.~\cite{Bao2023}. By choosing $J_{\perp} t / \hbar = \pi $, the initial state $\ket{\uparrow\uparrow}$ is converted to the maximally entangled Bell state $\frac{1}{\sqrt{2}}(\ket{\uparrow\uparrow} + i \ket{\downarrow\downarrow})$. The Bell state fidelity,  an important figure of merit for quantum computing, was found to be $\approx 0.8$ in both experiments, after correcting for state preparation and measurement errors. The fidelity is limited by motional dephasing of the spin-exchange oscillations, since the interaction energy  $\propto\langle r_{ij}^{-3} \rangle$  differs for each motional state, and the experiments were done using molecules in thermal states with a broad motional-state distribution. Sideband cooling to the ground state~\cite{Caldwell2020b,Lu2023,Bao2023sideband} will eliminate motional dephasing and also eliminate single-particle decoherence arising from non-uniform differential light shifts. 

\subsection{Scaling up}

Tweezer arrays can be scaled up to many hundreds of molecules, but this brings challenges. It will be necessary to control the tweezer separation with high precision and reproducibility throughout the array, maintain single-site addressability, avoid cross-talk and next-nearest neighbor interactions, and ensure the uniformity of the trap intensity and the addressing fields. In this context, it will be important to design protocols that are robust to inhomogeneous intensity and other non-uniformities.

\section{\label{Sec:outlook}Outlook and new directions}

\subsection{Controlling collisions and quantum degeneracy}

Understanding and controlling ultracold collisions is critical for studying molecular gases and  cooling them to quantum degeneracy. To date, experiments in optical traps have shown molecular collisions lead to loss for both reactive and non-reactive species~\cite{Ospelkaus:2010, Ni:2010,Takekoshi2014,Park2015,Ye:2018,Guo:2018,Gregory:2019,Yang2019}. Although not fully understood~\cite{Bause2023}, these losses are attributed to the transient formation of a collision complex~\cite{Mayle:2012,Mayle:2013} which subsequently absorbs photons from the optical trap leading to the observed loss~\cite{Christianen:2019,Gregory2019,Liu2020}. To circumvent this problem, several shielding methods~\cite{Langen2023} have been developed that exploit the dipolar interactions between molecules to engineer a repulsive barrier that prevents the molecules from reaching short range where the loss occurs. These methods include confining the molecules in 2D with their dipoles aligned perpendicular to the plane of confinement~\cite{deMiranda2011}, using a dc electric field tuned to a level crossing within the Stark-shifted rotational levels of the molecule pair~\cite{Avdeenkov2006} and dressing with strong microwave fields~\cite{Gorshkov2008,Karman2018,Lassabliere2018}. 

The dipolar interactions responsible for shielding collisions from loss can also increase the rate of elastic collisions. This is ideal for evaporative cooling, as spectacularly demonstrated with   KRb~\cite{Valtolina2020,Li2021} and NaK~\cite{Schindewolf2022}  to obtain the first quantum degenerate Fermi gases of molecules. Microwave shielding has recently been demonstrated for bosonic NaCs~\cite{Bigagli2023} and NaRb~\cite{Lin2023} molecules, resulting in the very recent first molecular Bose-Einstein condensates~\cite{bigagli2023observation}. This  opens new avenues, including  dipolar droplets and supersolids~\cite{Schmidt2022}, and provides a new route to the high-filling molecules in optical lattices for quantum simulation. Already it is evident that techniques to shield molecules from loss and to control their interactions will become an intrinsic feature of future experiments.

\begin{figure*}
    \centering
    \includegraphics[width=2\columnwidth]{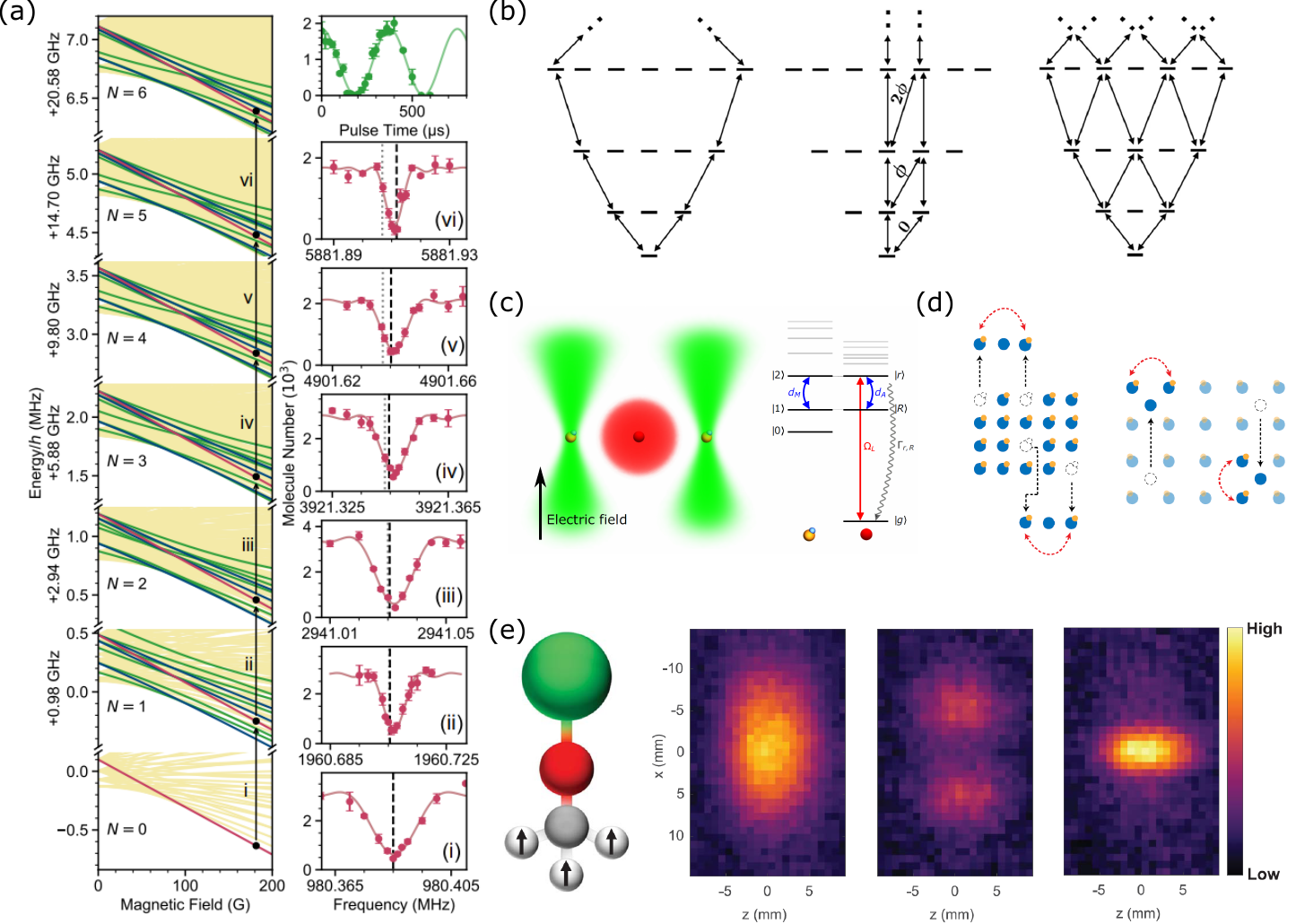}
    \caption{Future research directions. (a) Ultracold molecules offer a rich internal structure of long-lived rotational and hyperfine states easily accessible with microwave fields,   illustrated for $^{87}$Rb$^{133}$Cs up to $N=6$  prepared in the hyperfine ground state of $N=0$ at $\sim181$\,G. The microwave spectra (right) are labelled to match the transitions (left). The top panel shows Rabi oscillations on the $N=5\rightarrow6$ transition. The nuclear spins of $^{87}$Rb ($i=\frac{3}{2}$) and $^{133}$Cs ($i=\frac{7}{2}$) result in $32\times(2N+1)$ hyperfine states per  total angular momentum, $N$. (b) Rotational states can encode synthetic dimensions with the ``tunneling" rates controlled by microwave fields (arrows). Examples include (left to right) a 1D chain with periodic boundaries, a two-leg ladder with a gauge field and a square lattice with open boundaries. (c) \& (d) Interfacing molecules with Rydberg atoms may increase the effective interaction strength, allowing fast Rydberg-mediated  gates between molecules. Moving molecules and/or atoms using optical tweezers allows a scalable quantum computing architecture. (e) Cooling new  molecules will open new opportunities, such as the laser cooling of  CaOCH$_3$ symmetric-top molecules:  the unperturbed beam (left), Sisyphus heating (centre) and Sisyphus cooling (right).  (a) Adapted from~\cite{Blackmore2020}, (b) adapted from~\cite{Sundar2018}, (c) adapted from~\cite{ZhangTarbutt2022}, (d) adapted from \cite{Wang2022} and (e) adapted from~\cite{Mitra2020} , with permissions from the authors.}
    \label{fig:outlook}
\end{figure*}

\subsection{Beyond qubits: qudits and synthetic dimensions\label{sec:beyond-qubits}}

Molecules offer internal states across a vast range of energy scales: hyperfine, rotational, vibrational, and electronic. As Fig.~\ref{fig:outlook}(a) illustrates for RbCs,  hundreds of long-lived states are accessible with microwave fields in just a handful of rotational levels. Finding ways to harness this wealth of states is an emerging research area.

For quantum computing, Ref.~\cite{Sawant2020} proposed encoding qudits of dimension $d=4$ in the hyperfine states of a single rotational level of either RbCs or CaF molecules, and presented a protocol to implement the Deutsch algorithm. Ref.~\cite{Albert2020} considered quantum error-correcting codes encoding quantum information in  rotational-state superpositions, demonstrating theoretically robustness to relevant noise sources.

An exciting direction in quantum simulation is to harness the internal structure  to create \textit{synthetic dimensions}, in which motion in internal states is engineered to mimic motion in real spatial dimensions~\cite{ozawa2019topological,hazzard2023synthetic}. For example, sites of a lattice can be mapped onto the rotational states, with coherent ``tunneling" between sites controlled using microwave fields~\cite{Sundar2018}. The single-particle Hamiltonian is highly tunable: lattice geometry, tunneling rates (including gauge fields), and on-site potentials  can all be programmed (see Fig.~\ref{fig:outlook}(b) for examples). Moreover, with interactions, even the simplest one-dimensional synthetic dimension configuration is predicted to give rise to several phases of matter, and quantum  and thermal phase transitions~\cite{sundar:strings_2019,feng:quantum_2022,dasgupta:finite_2023}.

\subsection{\label{Sec:Rydberg}Increasing the interaction strength}

Although typical dipolar interaction strengths between molecules of $\sim0.1-10$\,kHz suffice for high-fidelity computation and simulation, access to stronger interactions is desirable. One way to achieve this is to use pairs of molecules in opposite spin states together with the state-dependent optical tweezers, in order to control the spacing of the molecules on a sub-wavelength scale. Ref.~\cite{Caldwell2020c} examined this, concluding that 2-3 orders of magnitude increase in interaction strength is feasible. 

An alternative approach is to use Rydberg atoms to mediate interactions between molecules~\cite{Kuznetsova2011,Kuznetsova2016,Zeppenfeld2017,Wang2022,ZhangTarbutt2022}. Rydberg atoms have transitions throughout the microwave domain, which  an electric or magnetic field can bring into resonance with a rotational transition in the molecule. The resulting resonant atom-molecule DDI can be much stronger than the interaction between molecules because of the large (typically $\sim$kD) transition dipole moment of the Rydberg atom. If the atom is placed between a pair of molecules, as illustrated in Fig.~\ref{fig:outlook}(c)\,\&\,(d), it can mediate a strong dipolar interaction between the pair. Two recent studies~\cite{Wang2022,ZhangTarbutt2022} have examined different protocols that exploit this mediated interaction to entangle  molecules. In both cases, a gate duration of $\sim1\,\mu$s and fidelity of 99.9\% look achievable. Very recently, the first observation of Rydberg blockade due to the charge-dipole interaction between an atom and a polar molecule has been reported~\cite{Guttridge2023}. The authors used species-specific optical tweezers to position a single RbCs molecule around 300\,nm   from a single Rb atom, sufficiently close to cause blockade of the transition to the Rb(52s) Rydberg state. Extensions of these experiments to perform non-destructive readout of the molecule~\cite{Kuznetsova2016,Zeppenfeld2017} and to engineer entanglement~\cite{Kuznetsova2011,Zeppenfeld2017,Wang2022,ZhangTarbutt2022} are eagerly anticipated.

\subsection{\label{Sec:richness}Different species}

To date, research into ultracold molecules has mostly focused on diatomic molecules. However, molecular diversity is enormous and attention is rightly turning to other species, including organic molecules and polyatomic species~\cite{Isaev2016}. Remarkably, the highly diagonal Franck-Condon factors needed for laser cooling can be found in a wide range of complex molecules. The common feature of these molecules is a ligand attached to an atom or diatomic species that acts as the optical cycling center. Already, magneto-optical trapping, sub-Doppler cooling and optical trapping of CaOH has been demonstrated~\cite{Vilas2022,Hallas2023} and many other species are under investigation (for a recent review see~\cite{Augenbraun2023}). In parallel, efforts are underway to extend the range of molecular species formed by associating ultracold atoms. Interspecies Feshbach resonances have been observed in several mixtures of alkali atoms with alkaline-earth-like atoms~\cite{Barbe2018,Green2020,Franzen2022}, although molecule formation has thus far proved elusive. In contrast, recent progress on $^{161}$Dy-$^{40}$K~\cite{Soave2023} and $^{6}$Li-$^{53}$Cr~\cite{Ciamei2022} mixtures looks promising. Evidence for the association of triatomic molecules in ultracold mixtures of $^{40}$K atoms and ground-state $^{23}$Na$^{40}$K molecules has been reported~\cite{Yang2022}. Recently, ``field-linked resonances'' have been observed in collisions between microwave-dressed $^{23}$Na$^{40}$K molecules~\cite{Chen2023}, opening up a route to coherent ``electroassociation'' of tetramer molecules~\cite{Quemener2023,Chen_arxiv}. 

Cooling new molecules to ultracold temperatures will allow new opportunities  stemming from their different structure and interactions with external fields. For example, polyatomic molecules exhibit  parity doublets where quantum states of opposite parity lie close in energy. This near degeneracy means that the states can be mixed with a small electric field, typically $\lesssim100$\,V/cm, leading to linear Stark shifts and making it exceptionally easy to produce laboratory frame dipole moments. These properties have stimulated a number of proposals advocating the use of symmetric-top molecules for quantum information processing~\cite{Wei2011,Zhang2017,Yu2019}, and laser cooling of CaOCH$_3$ has already been reported (Fig.~\ref{fig:outlook}(e))~\cite{Mitra2020}.

For quantum simulation, Ref.~\cite{Wall2013} highlighted the correspondence between the linear Stark shifts of a symmetric-top molecule and  a magnetic dipole in a magnetic field. They proposed that these molecules in  lattices could be used to simulate  spin models with arbitrary integer spin and in  later work that they may be used to study other quantum spin models, such as the XYZ  model~\cite{Wall2015}.

\section{\label{Sec:conclusion}Conclusion}

The field of ultracold molecules has seen rapid progress over the last decade. Through the combined effort of many research groups, the understanding and control of molecular systems has now reached a point where scientific applications beyond AMO physics are emerging. We have presented notable highlights in the realms of quantum simulation and quantum computation. However, this is undoubtedly just the beginning and we can expect many exciting developments in the near future. 
Looking further ahead, as the number of molecular species far exceeds the number of atomic elements, perhaps we should expect a day to come when molecules dominate the field of ultracold physics.

\begin{acknowledgments}
We acknowledge the support of UK Engineering and Physical Sciences Research Council (EPSRC) Grants EP/P01058X/1, EP/P008275/1, EP/V011499/1 and
EP/W00299X/1, UK Research and Innovation (UKRI) Frontier Research Grant EP/X023354/1, the Robert A. Welch Foundation (C-1872), the National Science Foundation (PHY1848304 and CMMI-2037545), the Office of Naval Research (N00014-20-1-2695 and  N00014-12-1-2665),  the W.F. Keck Foundation (Grant No. 995764), the Department of Energy (DE-SC0024301), and the Royal Society and Durham University.  
\end{acknowledgments}

\bibliography{refs}

\end{document}